%% file: main.tex
\documentclass[twocolumn,a4paper,10pt]{extarticle}

\usepackage{mathptmx}

\usepackage[compact]{titlesec}
\usepackage[hyphens]{url}
\usepackage{hyperref}
\usepackage[hyphenbreaks]{breakurl}

\titleformat{\section}[block]
{\fontsize{14}{15}\bfseries}
{\thesection}
{1em}
{}
\titleformat{\subsection}[block]
{\fontsize{11}{15}\bfseries}
{\thesubsection}
{1em}
{}

\usepackage[margin=0.78in]{geometry}

\usepackage{amsmath}
\usepackage{newfloat}
\usepackage{listings}
\usepackage{balance}

\usepackage[final]{microtype}
\usepackage[sort,nocompress]{cite}
\usepackage{adjustbox}

\usepackage{graphicx}

\usepackage{algorithmic}

\usepackage{booktabs,tabularx}

\usepackage{subcaption}

\usepackage{mathtools}
\usepackage{centernot}

\usepackage{cleveref}

\newcommand{\sam}[1]{\todo[author=Sam,color=cyan,inline]{#1}}

\renewcommand{\sam}[1]{}
\newcommand{\lev}[1]{\todo[author=Lev,inline]{#1}}
\renewcommand{\lev}[1]{}

\newcommand{\name}{Triangel}
\newcommand{\tochange}[1]{}
\newcommand{\ignore}[1]{}

\newcommand{\prerebuttal}[1]{\textcolor{black}{#1}}
\newcommand{\shepherd}[1]{{#1}}

\usepackage{cleveref}
\usepackage{framed} %
\usepackage{amssymb,amsmath,mathtools, amsthm}
\usepackage[framemethod=tikz]{mdframed}
\mdfdefinestyle{mystyle}{
	hidealllines=true,
	leftline=true,
	innerleftmargin=10pt,
	innerrightmargin=10pt,
	innertopmargin=0pt,
}

\newmdenv[
topline=false,
bottomline=false,
skipabove=\topsep,
skipbelow=\topsep
]{siderules}

\title{Triangel: A High-Performance, Accurate, Timely On-Chip Temporal Prefetcher}

\author{\hspace{2pt}Sam Ainsworth \hspace{110pt}  Lev Mukhanov \\  \hspace{20pt}  University of Edinburgh \hspace{40pt}  Queen Mary University of London \\  \textit{sam.ainsworth@ed.ac.uk} \hspace{60pt} \textit{l.mukhanov@qmul.ac.uk}
}

\date{}

\begin{document}

\maketitle
\newcommand{\speeduptriangel}{26.4\%}
\newcommand{\speeduptriage}{9.3\%}

\begin{abstract}

Temporal prefetching, where correlated pairs of addresses are logged and replayed on repeat accesses, has recently become viable in commercial designs. Arm's latest processors include Correlating Miss Chaining prefetchers, which store such patterns in a partition of the on-chip cache. 
However, the state-of-the-art on-chip temporal prefetcher in the literature, Triage, features some design inconsistencies and inaccuracies that pose challenges for practical implementation. We first examine and design fixes for these inconsistencies to produce an implementable baseline. We then introduce \name{}, a prefetcher that extends Triage with novel sampling-based methodologies to allow it to be aggressive and timely when the prefetcher is able to handle observed long-term patterns, and to avoid inaccurate prefetches when less able to do so. \name{} gives a \speeduptriangel{} speedup compared to a baseline system with a conventional stride prefetcher alone, compared with \speeduptriage{} for Triage at degree 1  \prerebuttal{ and 14.2\% at degree 4. At the same time \name{} only increases memory traffic by 10\% relative to baseline, versus 28.5\% for Triage}.

\end{abstract}

\input{sections/introduction}

\input{sections/triage}

\input{sections/inconsistencies}

\input{sections/triangel}

\input{sections/setup}

\input{sections/evaluation}

\input{sections/related}

\input{sections/conclusion}

\input{sections/artifact}

\balance
\small

\bibliographystyle{abbrv}

\end{document}

%% file: sections/introduction.tex
\section{Introduction}

The ever-growing memory wall has led to large classes of applications becoming latency-bound on cache misses. The solution in commercial cores has been to deploy an army of different types of prefetcher to bring in data to the caches before the user requests it, via prediction. Stride prefetchers~\cite{1194_Chen}, which predict incremental patterns in memory, and Spatial memory streaming~\cite{SMS-ISCA06}, which repeats patterns common to multiple regions of memory, have long been commonplace, but more recently, temporal prefetchers~\cite{Markov,GHB,ISB,Triage-MICRO19,TriageISR-ToC22}, which store and replay historical sequences, have been deployed in real cores~\cite{Arm-Hotchips}.

The Correlated Miss Chaining temporal prefetchers that have been deployed~\cite{Arm-Hotchips,X2-ECR,N2-ECR,A78-ECR} store the large volume of metadata necessary in a segment of the cache, and bear striking resemblance to Triage~\cite{Triage-MICRO19,TriageISR-ToC22}, a prefetcher published by authors from Arm and UT Austin. However, despite the recent commercial viability of temporal prefetchers, remarkably little exploration in the literature has occurred since. 

This paper seeks to ask two questions. First, is the design of Triage~\cite{Triage-MICRO19,TriageISR-ToC22} feasible for use in real processors? Second, are there simple improvements to the design that can significantly improve its timeliness, accuracy and/or coverage? Regarding the first question, our study suggests challenges, as certain properties of Triage are physically impossible to implement, while others are impractical due to design complexity or silicon area. Additionally, some design solutions result in severe performance degradation in edge cases.  %
We explore these to provide a foundation for a solid baseline.

\shepherd{For the second question, Triage~\cite{Triage-MICRO19} already observes that an aggressive high-degree prefetcher increases performance. We find performance can be improved further by increasing \textit{lookahead offsets} to store non-adjacent entries in the Markov table to improve timeliness, and by using metadata formats better able to deal with the physical-address fragmentation seen with a realistic operating system. Still, under its highest-performance configuration, Triage's accuracy drops to 50 percent~\cite{Triage-MICRO19}, which is unviable for energy-efficient cores.}

\shepherd{We introduce \name{}, a prefetcher based on Triage that adds several new structures to evaluate potential prefetches before we store metadata in the Markov table, to allow high aggression to still be efficient. We add new Samplers to observe long-term reuse and discover whether patterns will generate accurate prefetches. We add a Metadata Reuse Buffer, to allow high-degree aggressive prefetching without increasing traffic to the L3 cache, where prefetch metadata is stored. We also replace Triage's Bloom-filter sizing mechanism with a novel Set-Dueller, which is able to model arbitrary cache-partition configurations to find the best tradeoff between prefetch metadata and cache hit rates. } %

\name{} gives \speeduptriangel{} geomean speedup compared to a baseline with stride prefetcher alone, versus \speeduptriage{} for Triage, on the same workloads and similar core setup to the original papers~\cite{Triage-MICRO19,TriageISR-ToC22}. \prerebuttal{It does this while being significantly more efficient, at only 10\% memory-traffic increase versus 28.5\% for Triage.} \prerebuttal{When Triage is run more aggressively to counteract its baseline's low performance, by being unconditionally degree-4 (\name's maximum degree), it still only achieves 14.2\% geomean speedup with 43.8\% memory traffic overhead, meaning Triangel represents a Pareto improvement in both efficiency and performance against all Triage configurations.}

%% file: sections/triage.tex
\section{Background: Triage}

Triage~\cite{Triage-MICRO19} is a temporal prefetcher published by authors from UT Austin and Arm in 2019. It is an address-correlating Markov prefetcher~\cite{Markov}, in that it stores $(x,y)$ address pairs: when an L2 cache miss (or tagged prefetch hit, when data that was prefetched into the cache is first used) to $x$ occurs, $y$ is prefetched, as the recorded miss (or prefetch hit) that occurred after $x$ the last time it was brought into the cache.

\begin{figure}[t]
    \centering
    \includegraphics[width=\columnwidth]{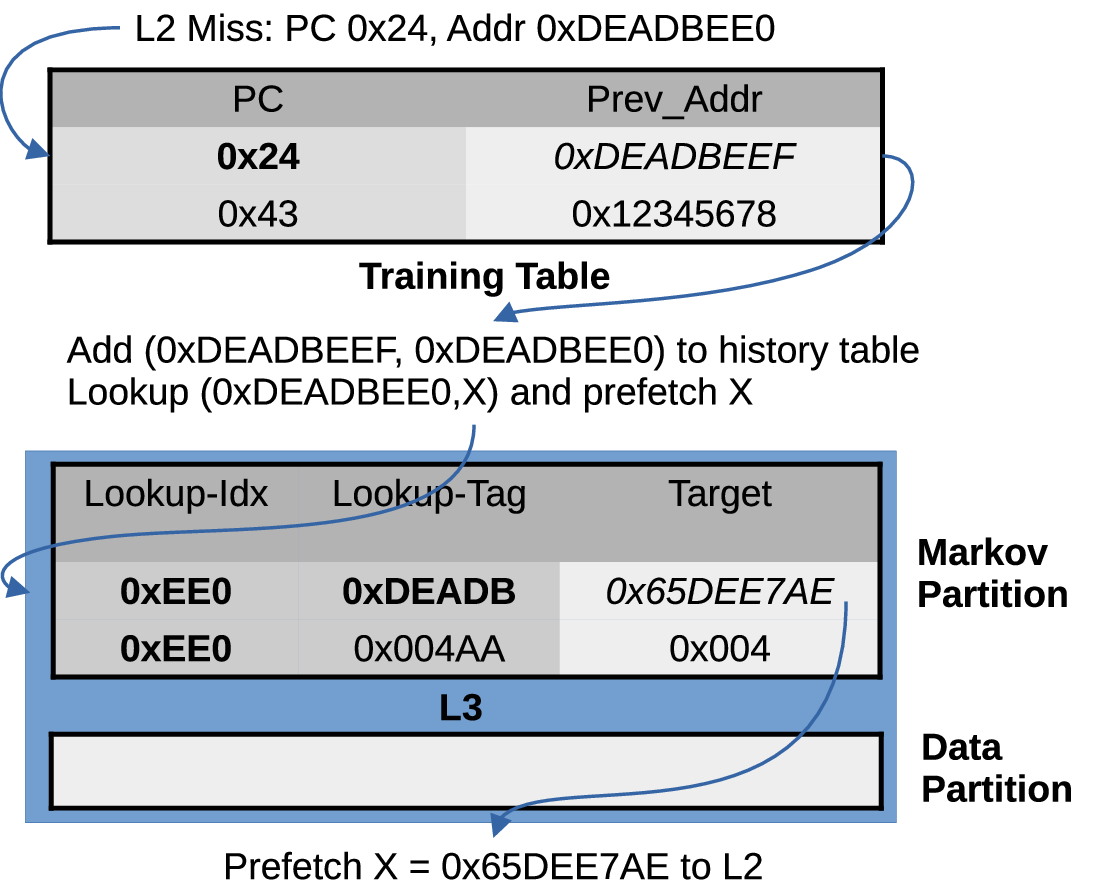}
    \caption{The basic operation of Triage~\cite{Triage-MICRO19,TriageISR-ToC22}. On a cache miss (or tagged prefetch hit), the PC is used to index the training table. The previous address is used as an index to train the Markov history table. The current access is then looked up in the Markov table to generate a prefetch. Not shown: index- and target-compression mechanisms (\cref{ssec:metadataformat}), confidence bits (\cref{ssec:confidence}), partition sizing (\cref{ssec:triagesizing}).}
    \label{fig:triage}
\end{figure}

Triage is PC-localized, in that it separates patterns based on the PC that accesses the location. This manifests as a training table indexed by PC, used to store the previous miss/tagged hit for each PC.
This training data then feeds into a Markov table (storing historically correlated address-pairs from the miss/prefetch-hit stream) that is not tagged by PC, and is stored as a variable-sized partition of the L3 cache. A detailed example of its behavior is given in \cref{fig:triage}. %

Triage-ISR~\cite{TriageISR-ToC22}, published in 2022, is an extension by the same authors. It changes the method of Markov-table compression (\cref{ssec:metadataformat}), simplifies the Markov-partition sizing mechanism (\cref{ssec:triagesizing}), and extends the Markov-table format to represent groups of contiguous locations\footnote{This extended format gives only a marginal speedup, and reduces capacity for non-sequential accesses to $3/4$ of the original, extending entries from 32 bits to 42 bits, but achieving only 3\% compression. We leave it as orthogonal for this work, though we consider Triage-ISR's other features in \name{}.}.

Triage has influenced temporal prefetchers in production. Arm's Correlated Miss Chaining~\cite{Arm-Hotchips} prefetcher, introduced in the X1 and A78 onwards~\cite{Anand-A78X1}, stores and replays Markov-table pairs in a partition of the L2 cache~\cite{A78-ECR,N2-ECR}. The Cortex X4~\cite{WikichipX4} adds another temporal prefetcher at the L1 cache.

%% file: sections/inconsistencies.tex
\section{Fixing Inconsistencies in Triage}
\label{sec:inconsistencies}

Although the original papers present an intriguing approach, many of the details in these papers~\cite{Triage-MICRO19,TriageISR-ToC22} are missing, incomplete, and/or impossible to implement.
Here we work through the subtler details. Where it is implementable, we use the most modern mechanism at the source of our baseline~\cite{TriageISR-ToC22}, and refer to this as ``Triage'' in our evaluation. Where ambiguity is impossible to clear up otherwise, we reference a public implementation contributed to by an author of the original papers~\cite{Triage-Codebase}, which does not always match either paper but sheds light on some inconsistencies and design choices. %

 We first cover fixes for these, choose the most appropriate option for our baseline where no one solution is obvious, and discuss inefficiencies, before discussing how to design an energy-efficient, timely, accurate and high-coverage temporal prefetcher, \name{}, in the following section.

\begin{figure}
\centering
\subfloat[\prerebuttal{Lookup-Address Example: 0xDEADBEEF looks at the 16 entries inside cache index 1, and finds a match of its hashed tag at the blue element.}] {
 \includegraphics[width=\columnwidth]{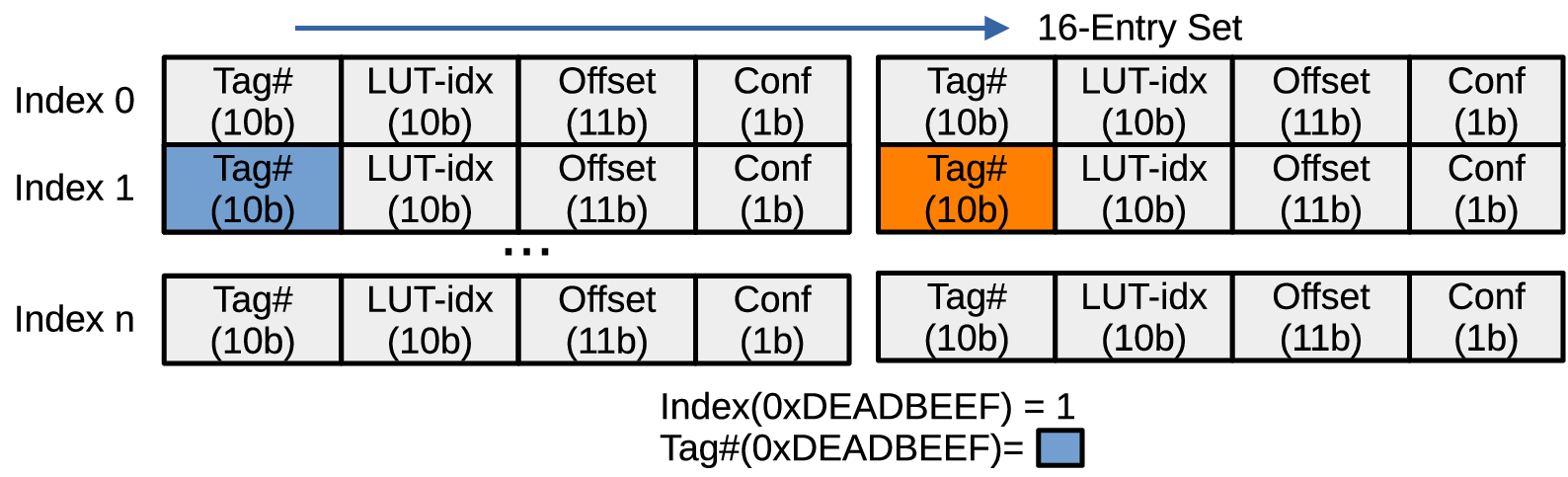}
 \label{fig:triagelookup}
}

\subfloat[\prerebuttal{Prefetch-Target Example: the lookup-table index (LUT-idx) indirects into entry 64 of the lookup table, and the result is combined with the 11-bit offset flag and 6 zero-bits for the cache line, to create a full address.}]{
 \includegraphics[width=\columnwidth]{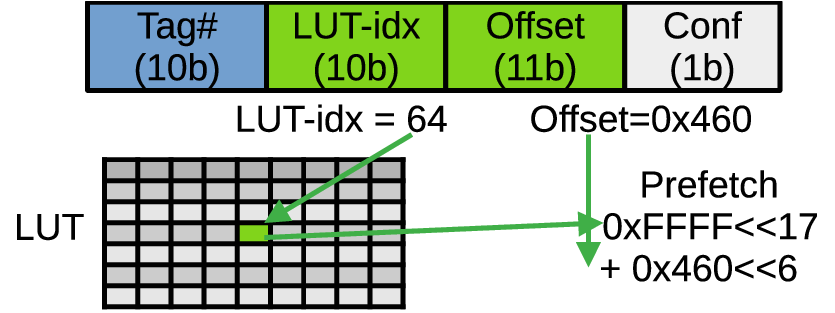}
}
    \caption{Fields in the Markov table~\cite{Markov} segment of the cache in our reimplementation of Triage~\cite{Triage-MICRO19,TriageISR-ToC22}. Its lookup address is indexed by cache set and sub-set (\cref{ssec:index}, and tagged by an XOR hash of the full address (tag-\#). The prefetch target is generated by using the LUT-idx bits as an index into the 1024-entry lookup table, which is then combined with the Offset and 6 zero bits for cache-line alignment.}
    \label{table:mhistory}
\end{figure}

\subsection{Markov Metadata Format}
\label{ssec:metadataformat}

Triage~\cite{Triage-MICRO19} and Triage-ISR~\cite{TriageISR-ToC22} have very different approaches to storing Markov-table metadata (temporally correlated pairs of addresses). While Triage-ISR fixes some of the implementation complexities in Triage, it introduces other mechanisms that we believe are impossible to implement. The table we use for our Triage experiments is based on choosing the implementable parts of each, and is shown in \cref{table:mhistory}.  %

In both original techniques, two addresses (the \textit{lookup address} and \textit{prefetch target}) are compressed to fit both within 32 bits total. Each address is cache-line aligned, so the 6 least significant bits are implicit (always zero). 
We assume addresses are physical, typically without loss of generality.

In Triage~\cite{Triage-MICRO19}, both addresses are treated identically. 11 bits of the \textit{lookup address} are stored implicitly based on the set that is indexed during its lookup. For the \textit{prefetch target}, these 11 bits are stored explicitly in the 32-bit Markov metadata entry. The remaining ``tag'' bits for each are stored as 10-bit indices into a (presumably 1024-entry) lookup table. %
\prerebuttal{Using the same lookup table for both causes issues, in that the index of the L3 cache must be exactly 11 bits as a result (2MiB for a 64-byte-line 16-way set-associative cache), otherwise the offset for the otherwise unrelated prefetch target inside the 32-bit Markov-table entry must grow, meaning the Markov-table entry \textit{itself} must grow to unaligned sizes to still fit both entries.} %

This lookup table is likely stored in a separate SRAM structure rather than a cache partition: e.g. 4-byte tags would result in only a 4KiB structure. Its capacity is limited by the number of available index bits in the Markov table's fields. 
\prerebuttal{It could be considered a simple array looked up via index, except} finding the correct 10-bit tag for a lookup address, to access the correct Markov-table entry, requires a reverse lookup in the same table, as does checking for an existing mapping on inserting a new prefetch target.  \prerebuttal{This structure must therefore also support cache-like indexing; we found that implementing lookups via 16-way associative indexing and replacement performed similarly to fully associative (\cref{ssec:triageanalysis}).} %

Likely due to the above issues, Triage-ISR~\cite{TriageISR-ToC22} removes this table entirely. \shepherd{For the lookup address}, a 7-bit hash is stored instead of the full tag being retrieved via lookup table. However, the solution they give for the prefetch target is impossible to implement. They suggest storing the address as a 16-bit index of the L3 cache, followed by a 7-bit hash of the remaining bits.  This is fine if the address to be prefetched is still in the L3 cache, because cache presence can act as a heuristic to choose candidates: any addresses that match the index and tag-\# that are currently in the L3 can be brought into the L2\footnote{Neither Triage's nor Triage-ISR's mechanism is implemented in the public code~\cite{TriageISR-ToC22}. %
Triage-ISR's strategy may be valid in cases where the Markov table is stored in the L2, as is the case in CMC~\cite{X2-ECR,Arm-Hotchips}, since the Markov table's range is less able to exceed the capacity of the L3.}. However, it is impossible to identify a target to be brought in from main memory with only these 29 bits (with 6-bit line offset, 16-bit cache index, and 7-bit hash). It can only represent 512MB of unique addresses: all the many addresses that match in DRAM are equally likely to be the valid target. In essence, to use it involves inverting a many-to-one hash.

It is possible to generate a working technique by combining the two (\cref{table:mhistory}). The lookup address should be stored as in Triage-ISR, with an implicit index combined with a 10-bit hashed tag\footnote{With 16 elements in each cache line, and 8 ways in the Markov partition, there are 128 possible candidates. 7 bits is insufficient and reduces performance; the probability of collision for each new insertion is 0.634 according to a binomial distribution. We increase the hashed tag size to 10 bits as a result because of the resulting performance loss, both for Triage and Triangel.}. \prerebuttal{This avoids the cache requiring any particular indexing policy, as the table-lookup for the prefetch target is completely decoupled. For the lookup address, the hash of the remaining bits of the address not in the index will be fixed-size regardless of how many bits are in the index itself.}

We find that  even slight changes in the workload's frame locality cause severe slowdown (\cref{ssec:triageanalysis}), %
 and so \name{} stores the physical address directly in the Markov table.%

\subsection{Associativity and Indexing}\label{ssec:index}

Triage and Triage-ISR store Markov-table entries~\cite{Markov}, compressed inside cache lines of the L3 cache, with Triage storing 16 entries per 64-byte cache line, each containing its own independent tag stored within the cache line's data, rather than the single tag for the cache line.  This raises questions on how indexing works, and how set-associative the structure ends up being, that are not answered in either paper~\cite{Triage-MICRO19,TriageISR-ToC22}.

Finding a Markov-table element requires fetching each cache line's data rather than just the tags, and up to eight ways are allocated to the Markov partition. It would take 160 cycles to access all possible tags of the resulting 128-way associative structure as a result, rather than the 20 cycles in the paper.   %

The public codebase's~\cite{Triage-Codebase} Markov table is set to be either 0, 4 or 8-way set associative, depending on whether the metadata partition of the history table uses 0, 4 or 8 ways of the 16-way last-level cache. Presumably, the 16 elements inside each cache line are indexed as direct mapped, with each index only storable in one location within a cache line, but in any cache line in the set. This is impractical: it would still take 160 cycles to access just one compressed tag from within each individual line, since the Markov tags are inside the cache lines. %

Our policy is always 16-way associative, accessing just one line, by using a second indexing policy, in addition to the bits choosing the cache set. This chooses the \textit{sub-set}: the relevant cache line/way within the L3 cache's set. We take the modulus of the 10-bit tag-\#, where $Partition\_Ways$ is the current number of ways reserved for Markov metadata:
\begin{center}
$Index = Tag\text{-}\#\ \%\ Partition\_Ways$
\end{center}
The sub-set index for a given address changes every time the partition size changes, so previously filled Markov-table entries end up inaccessible due to being in the wrong sub-set. We trigger a rearrangement by storing the current indexing policy (3 bits) of a given set in the first cache line's tag bits (which are otherwise unused).
On an access, if this does not match the current partition size, the entire set is rearranged asynchronously following the access.   %

The original papers also give no explicit information on where LRU or HawkEye replacement state is stored. LRU state can be stored implicitly within a line by storing the most recently used element at index 0 and shifting down other elements -- PLRU~\cite{rmPLRU,pseudolru} bits or RRIP~\cite{RRIP} bits can be stored within remaining unused cache-line tag bits.

\subsection{Cache Replacement}
\label{ssec:hawkeye}

Triage and Triage-ISR use HawkEye~\cite{HawkEye}, a complex cache-replacement technique, to prioritize more frequently used Markov-table entries when space-constrained. This stores a long history of 64 randomly chosen sets (set duelling~\cite{SetDuel}) to classify PCs based on whether Belady's optimal algorithm would choose to store data brought in by them in the cache, performing an $O(n)$ walk of the data structure on each access. Negatively classified PCs insert metadata in the most evictable way, and never promoted above positively classified PCs' loads, whereas postively classified PCs are directly inserted with a high reuse (to deter eviction) and treated as LRU.

With a default 1MiB-maximum Markov table, we found a relative speedup of only 0.25 percent from HawkEye over LRU for the 7 SPEC workloads in the original papers. The benefit became noticeable only with an artificial limit for the prefetcher of 256KiB maximum state\footnote{We saw even less benefit for Triangel, which already performs filtering of entries before cache replacement. Artificially limited to 256KiB of state, Triangel saw 2.8\% geomean for LRU, 4.3\% for RRIP and 6.1\% for HawkEye, versus Triage's 1.2\% for LRU, 2.9\% for RRIP and 5.8\% for HawkEye.}.  The storage and use of Markov-table entries is not the same problem as cache replacement. To consider whether a prefetch entry should be stored, we should not only consider if it is accessed (as with a cache line). We also need to consider whether the prefetch will be useful. This only occurs if there is some form of (semi-) sequential pattern to accesses for a given PC. %

\subsection{Confidence Bit}
\label{ssec:confidence}

One bit in the Markov table is used as \textit{``a confidence counter"}~\cite{Triage-MICRO19}, with no other information provided. In the public implementation~\cite{Triage-Codebase}, this is just used for same-index replacement: If $(x,y)$ is stored in the table, and we see $(x,z)$, then replacement only occurs if the confidence bit is not set. We follow the same design; the alternative would be using the bit to guide prefetching itself, not prefetching unless the confidence bit is set. However, we discovered experimentally that this was too pessimistic, and thus hurt performance. \tochange{In Triangel, where the utility of a prefetch can be predicted more accurately, we use this bit more selectively for PCs with only semi-predictable usage patterns (\cref{ssec:aggression}).}

\begin{figure*}
    \centering
    \includegraphics[width=1.7\columnwidth]{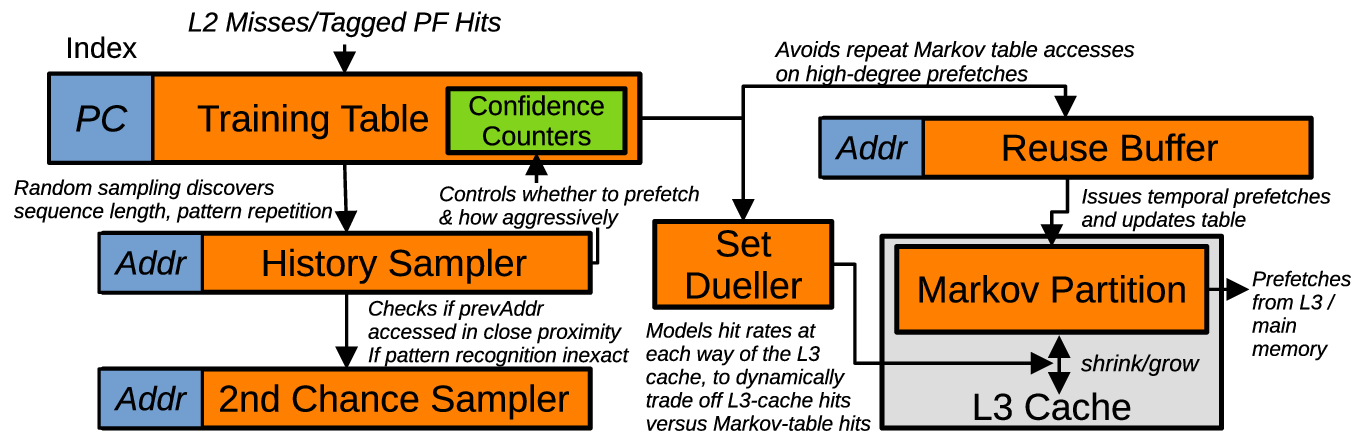}
    \caption{The structure of \name{}. Like Triage, it tracks per-PC miss sequences $(x,y)$ in the training table, and stores and replays them using a Markov table~\cite{Markov} inside a partition of the L3 cache. \name{} adds four new structures: a History Sampler, which randomly samples the training table to observe long-term patterns, a Second-Chance Sampler to identify inexact sequences that still give accurate prefetches, a Metadata Reuse Buffer to eliminate duplicate L3 Markov-partition accesses from high-degree prefetches, \prerebuttal{and a Set Dueller, to choose the partitioning of L3-data-cache versus Markov table that optimizes hit rates.}}
    \label{fig:triangel}
\end{figure*}

\subsection{Table Sizing}
\label{ssec:triagesizing}

To choose the number of L3 cache ways that form the Markov-table partition, Triage-ISR~\cite{TriageISR-ToC22} uses a Bloom filter~\cite{10.1145/362686.362692} trained on every access to the prefetcher within a 30-million-instruction window: if the address is a bloom-filter miss, then it has not been seen before, and so the target size of the partition is increased to fit it. This allows more fine-grained set-allocation decisions than the binary 0, 4, 8 from the original Triage~\cite{Triage-MICRO19}, which was set by comparing the results of two HawkEye predictors, and which incurs significant implementation complexity~\cite{TriageISR-ToC22}. Likewise, we can infer from Arm's documentation of CMC~\cite{A78-ECR,X2-ECR} that it too supports fine-grained partitioning. We use the Bloom-filter design for our Triage baseline. The exact sizing, as with all other structures in Triage, of the bloom filter is not given. But even a Bloom filter with 5 percent chance of false positives is 200KiB~\cite{bloomsize}; too large to be at the side of each L2 cache. The Triage-ISR Bloom-filter strategy also wastes space: there is a persistent bias towards Markov-table entries regardless of how useful the L3 cache would otherwise be: if there are unique Markov-table indices that could be stored, the partition will grow to fit them regardless of their utility or the effect on DRAM traffic from the reduced L3 capacity. This is true regardless of whether HawkEye would prioritize them, in that even entries judged as non-temporal are added to the Bloom filter\footnote{In HawkEye~\cite{HawkEye}, the extra space fills up with old, obsolete Markov entries rather than newer non-temporal entries, as a non-temporal entry is always evicted in preference to a temporal entry, and temporal entries do not age beyond non-temporal entries (unlike in RRIP~\cite{RRIP} or Mockingjay~\cite{Mockingjay}).}.

%% file: sections/triangel.tex
\section{\name}

Now we have an implementable baseline for Triage, we wish to improve upon its energy efficiency, performance, accuracy and timeliness. The broad strategy is as follows:

\begin{itemize}
    \item We use sampling techniques to estimate whether a PC is likely to generate good prefetches, filtering away the rest from being stored or used (\cref{ssec:historysampler}).

\item We improve timeliness and performance by increasing prefetch aggression when we are confident (\cref{ssec:aggression}).

\item We lower energy use by eliminating poor-quality prefetches (\cref{ssec:aggression})\prerebuttal{, filtering repeat L3 Markov-table accesses from high prefetch degrees (\cref{ssec:metadatareusebuffer}), and by dynamically trading off Markov-table versus L3-cache hit rates (\cref{ssec:metadatasize}}) to mitigate DRAM traffic. %

\end{itemize}

\subsection{Basic Operation}

The basic structure of \name{} is given in \cref{fig:triangel}. Like Triage~\cite{Triage-MICRO19,TriageISR-ToC22}, it stores a training table to track misses/tagged-prefetch hits per-PC, with new fields and counters  (\cref{ssec:training}) to support \name{}'s aggression control.

\begin{figure}
    \centering
    \includegraphics[width=\columnwidth]{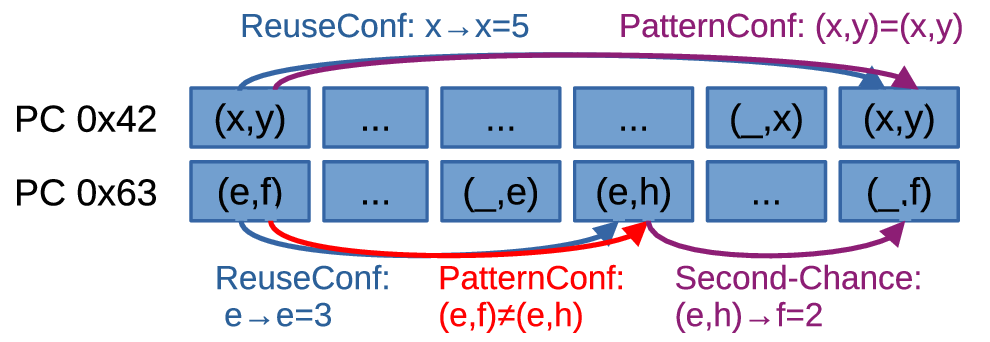}
    
    \caption{An example of the classifications performed by Triangel's samplers. ``\_'' signifies an arbitrary address. For PC 0x42, sampling \textit{(x.y)} reveals that $x$ is repeated within a region short enough to be stored in our Markov table (ReuseConf). Since \textit{y} is also accessed following \textit{x} on \textit{x}'s repeat, the pattern repeats (PatternConf) and thus temporal prefetching is accurate. For 0x63, when \textit{e} repeats, it is followed by \textit{h} rather than the \textit{f} we expect. However, Second-Chance Sampling (\cref{sssec:secondchance}) reveals that we access \textit{f} nearby, and so a prefetch to \textit{f} at \textit{(\_,e)} would be used before eviction, despite the imperfect sequence.     \textit{Note the Markov table can only store one target per index, so will store only one of (e,f) or (e,h) at any given point.}}  %

    \label{fig:classifiers}
    
\end{figure}

\name{} uses per-PC random sampling to evaluate the likelihood of a previous $(x,y)$ sequence being likely to generate an \textit{accurate} future prefetch (\textit{PatternConf}) before it is \textit{evicted} from the Markov table (\textit{ReuseConf}). Sequence lengths and exact pattern repetitions are analyzed by the History Sampler (\cref{ssec:historysampler}). If the History Sampler finds a different target $z$ for a repeat index $x$ that previously targeted $y$, it defers judgement to the Second-Chance Sampler (\cref{sssec:secondchance}), \prerebuttal{which checks if previous successor $y$ is accessed in close proximity (causing a hypothetical prefetch to $y$ to still be accurate). Examples are given in \cref{fig:classifiers}.}

These confidence counters control prefetcher aggression (\cref{ssec:aggression}): not only whether prefetches and Markov-table updates occur (to improve accuracy/energy efficiency), but also how far ahead to prefetch: if the classifiers indicate high confidence, lookahead distance and degree are increased. High-degree prefetching is made energy efficient and timely by the addition of a Metadata Reuse Buffer (\cref{ssec:metadatareusebuffer}): if chained walks through multiple future table entries are saved locally, when the sequence is re-walked for following prefetches, we avoid redundant L3 accesses.
\prerebuttal{Finally, the Bloom filter is replaced by a Set Dueller that directly trades off L3 data-cache storage and Markov-table storage (\cref{ssec:metadatasize}).}

\subsection{Training Table}
\label{ssec:training}

\begin{figure*}[t]
    \centering
  \begin{adjustbox}{max width=\textwidth}
  \begin{tabular}{|c|c|c|c|c|c|c|c|}
  \hline
        PC-Tag-\# & LastAddr$[0]$ & \textbf{LastAddr$[1]$} & \textbf{Timestamp} &  ReuseConf & \textbf{PatternConf} & \textbf{SampleRate} & \textbf{Lookahead}\\
       10 bits & 31 bits & 31 bits & 32 bits  & 4 bits & 2x4 bits & 4 bits  & 1 bit\\
        \hline
    \end{tabular}
    \end{adjustbox}
    \caption{Fields in the training table, which is indexed and tagged by PC. Bold fields are new to Triangel, others are taken from Triage (assuming Triage uses a saturating counter of the same size as ReuseConf for HawkEye classification~\cite{HawkEye}).}
    
    \label{tab:training}
\end{figure*}

We add several fields to Triage's training table for \name{}'s filtering and aggression analysis. It is indexed by PC, and is updated on a miss or tagged prefetch hit (when a prefetched element is accessed for the first time, so would have missed without prefetching) in the L2. %
Its layout is given in \cref{tab:training}:

\begin{itemize}
    \item \textbf{PC-Tag-\#}: Hashed tag of the PC (similar to Triage-ISR's hashed tags, \cref{ssec:triageanalysis}), to identify entries in the table.
    \item \textbf{LastAddr$[0,1]$}: This stores the previous misses/prefetch hits observed at this PC, as a shift register. It is used in the History Sampler, and in the Markov table if the History Sampler (\cref{ssec:historysampler}) positively classifies the PC. When LastAddr$[0]$ is filled, its previous value is shifted into LastAddr$[1]$. The latter is used as the Markov-table index instead of the former when the prefetcher is in an aggressive state, increasing lookahead (\cref{ssec:aggression}).
    \item \textbf{Timestamp}: This is a per-PC local timestamp, incremented every time the training-table entry is accessed. It calculates distance between repetitions (\cref{ssec:historysampler}). %
    \item \textbf{ReuseConfidence}: A saturating counter to evaluate whether the address pattern at PC $x$ repeats in a short enough sequence to fit in the Markov table (\cref{ssec:historysampler}). 
    \item \textbf{PatternConfidence} \prerebuttal{Two 4-bit saturating counters, biased by different factors, to evaluate prefetching accuracy: they consider if, for an $(x,y)$ stored pair, address $y$ is likely to be accessed in proximity to address $x$ the next time $x$ is accessed (\cref{ssec:historysampler}). The first counter saturates if we are $>66\%$ confident (enough to store the metadata and issue one prefetch from it) and the second to saturate if we are $>83\%$ certain (to issue high-degree prefetches).}
    \item \textbf{SampleRate}: Controls sampling rate for a PC, to balance observation frequency with being able to observe repeat accesses before eviction from the sampler (\cref{ssec:historysampler}). %
    \item \textbf{Lookahead}: Stores whether we currently use LastAddr$[0]$ versus LastAddr$[1]$ as the index for Markov-table training, for aggression control via lookahead (\cref{ssec:aggression}). 
    
\end{itemize}

\begin{figure}[t]
    \centering
\includegraphics[width=.7\columnwidth]
{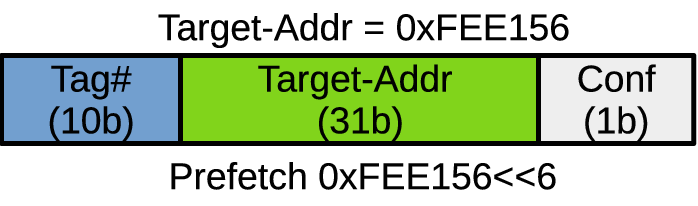}
    \caption{\prerebuttal{Markov-table fields in \name, and example prefetch-target generation. Its lookup-address process is based on our Triage reimplementation (\cref{fig:triagelookup}). The prefetch target does not use a lookup table, instead being} generated by shifting the 31-bit \textit{target-addr} with 6 cache-line zero bits, with 128GB range.}
    \label{table:history}
\end{figure}

\subsection{Markov Table}
\label{ssec:markov}

The Markov table (\cref{table:history}) is placed in a segment of the L3 cache. It stores pairs of addresses: one to address the table (\textit{lookup address}), and the other a prefetch (\textit{prefetch target}).
Triangel's table stores 12 compressed elements inside each 64-byte cache line, and like Triage-ISR~\cite{TriageISR-ToC22}, uses a (10-bit) lookup-address hash instead of a full tag. Like Triage and Triage-ISR, we store a single \textit{confidence} bit for replacing the \textit{prefetch target} for a given lookup address: a target is replaced if \textit{confidence} is 0, which is set to 1 if the new prefetch target on training matches the existing prefetch target in the table.

\subsection{History Sampler}
\label{ssec:historysampler}

The History Sampler is used to make decisions about whether to store history for a given PC. It does this by taking samples\footnote{Simple methods such as linear congruential~\cite{10.1145/76738.76798} are fine; cryptographic randomness is not required. The important factor is that sample rate can be varied to read more or less of the address stream.} of the metadata in the training table, such that it can see much further into the past than the data stored in the cache itself, despite being small and 2-way associative. %

\begin{figure}[t]
    \centering
 \includegraphics[width=.9\columnwidth]{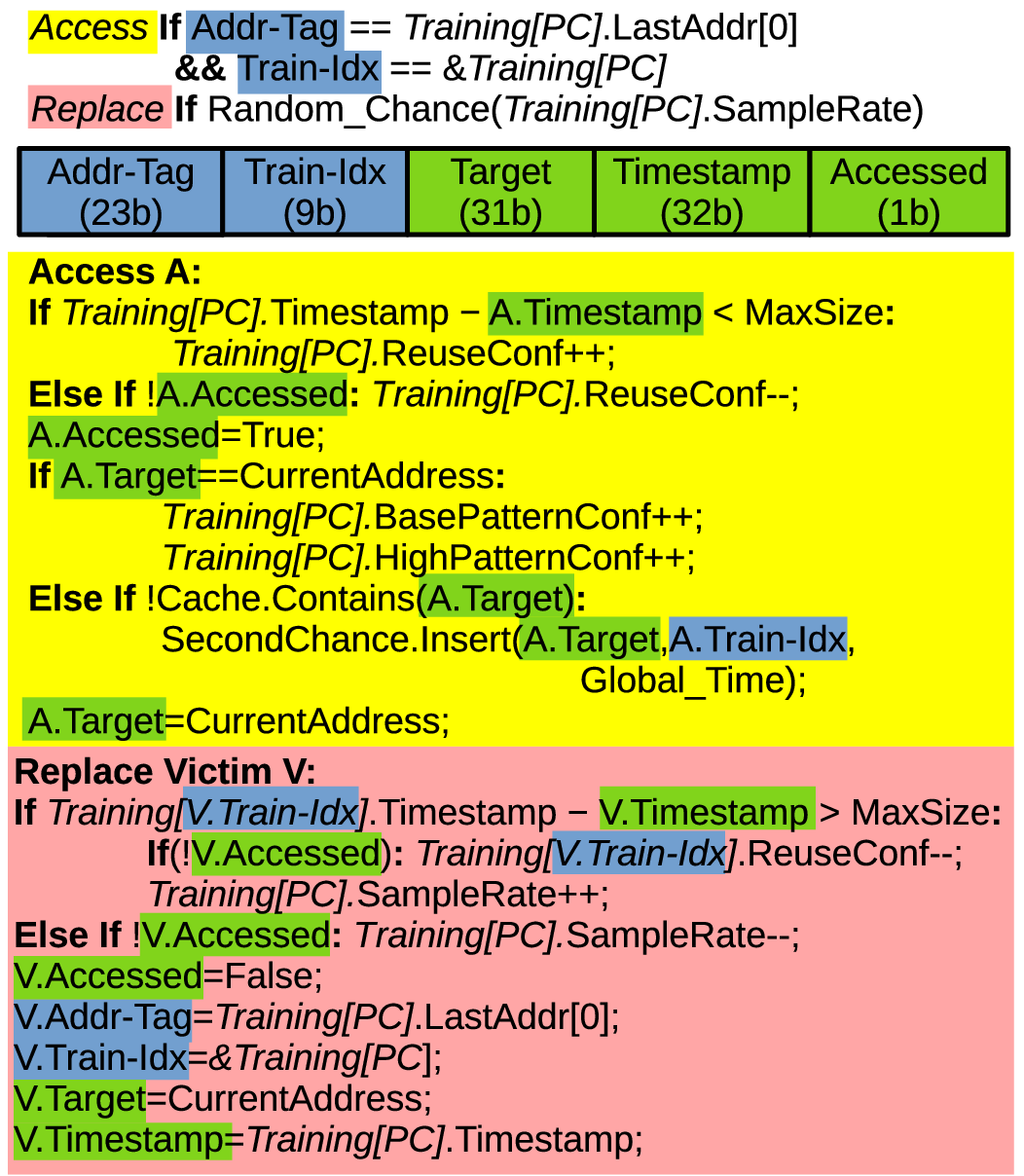}
    \caption{\prerebuttal{Layout and pseudocode for the History Sampler. }}
    \label{tab:sampler}
\end{figure}

The entries stored and pseudocode for the History Sampler are given in \cref{tab:sampler}. Every time the training table is updated (L2 cache miss or prefetch hit), LastAddr$[0]$ is looked up in the sampler along with a check that Train-Idx matches the index of the current PC's training-table entry. If we see a hit, and the difference in timestamps (between the one in the training-table entry and in the sampler) is below a threshold, we consider the pattern small enough to store in the L3's Markov-table partition (\textit{ReuseConf}). If the CurrentAddress being trained on by the prefetcher matches the Target in the sampler entry, and the Train-Idx matches the training-table entry for the current PC, we consider the prefetch to be accurate (\textit{PatternConf}).

\subsubsection{Reuse Confidence}

If the looked-up address is in the History Sampler, and the table index stored matches the training-table entry for the current PC, we have a repetition of a memory access we have seen before. We next calculate its local reuse distance to evaluate whether, ignoring the existence of any other PCs, the pattern for this PC alone is short enough to be prefetched within the Markov Table's \prerebuttal{\textit{MaxSize} (196608 entries for a 1MiB partition of 42-bit entries)}.
The local timestamp in the training table is incremented each time the individual (per-PC) training-table entry is accessed, and when an entry is sampled, the timestamp is copied into the History Sampler. We subtract the two on a hit to find their distance. If this is below \prerebuttal{\textit{MaxSize}, we increase ReuseConf\footnote{\prerebuttal{Where multiple PCs together fetch more than the \textit{MaxSize} but not individually, we found it was best to leave the decision of which to prefetch to cache replacement in the Markov table, and whether to prefetch at all to the Set Dueller, rather than try to choose based on reuse distance which ones win or lose. In this sense, ReuseConfidence is a weak classifier that only removes entries that are always useless; we hope future heuristics will be more precise.}}}.

\subsubsection{Pattern Confidence and Second-Chance Sampler}
\label{sssec:secondchance}

Unlike for cache replacement, prefetcher data being re-accessed alone is insufficient to make a prefetch useful. If the pattern of memory accesses is unpredictable, we will both pollute the cache and needlessly increase DRAM traffic.

We track if the \textit{pattern} $(x,y)$ repeats, instead of just an individual access $x$. If the PC's training-table entry's $lastAddr[0]$ is $x$, $(x,y)$ is in the sampler, and the $currentAddress$ just seen is also $y$, we can increase PatternConf. 

\begin{figure}[t]
    \centering
\includegraphics[width=\columnwidth]{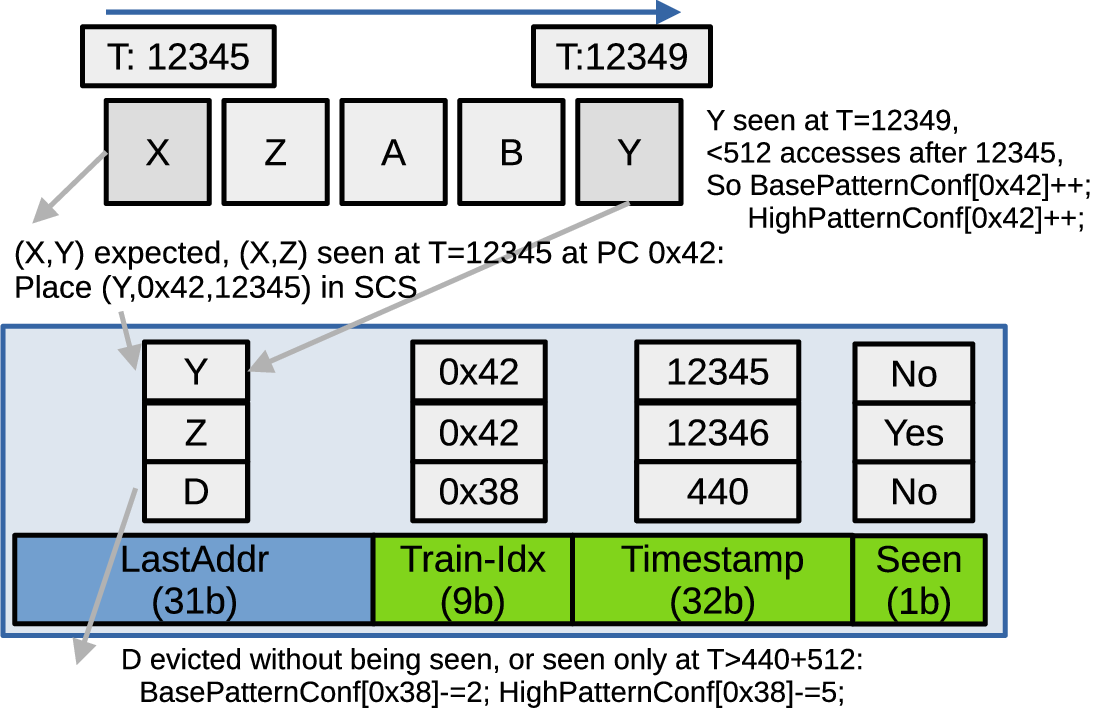}
    \caption{\prerebuttal{Second-Chance Sampling finds more general temporal correlations than the immediate sequences covered by the History Sampler. If \textit{(x,y)} are collocated in the History Sampler but not the Training Table, we work out whether a prefetch to y would be accurate despite the mismatch by storing y and a timestamp  in a small buffer. If y is accessed within 512 training accesses, we increment confidence and otherwise decrement it.}}
    \label{tab:secchance}
\end{figure}

We decrement PatternConf if $currentAddress$ does not match a sampled entry for $lastAddr[0]$,  with two exceptions. \prerebuttal{If the sampler's target is already in the cache} %
(so would not generate a prefetch, inaccurate or otherwise) we leave counters at their old values.
To catch when a pattern is not a perfect sequence, but the hypothetical prefetch to the target $y$ in the history sampler would still be used before cache eviction (so is  an accurate prefetch), we add a small table: the Second-Chance Sampler (SCS, \cref{tab:secchance}). If a target in the History Sampler does not match the currentAddress, it is placed in the SCS. When the training table is updated, the SCS is checked in addition to the History Sampler. If the currentAddress is in the SCS, the PC matches, \prerebuttal{and we are within 512 fills to the L2 cache (an underapproximation of L2 capacity), PatternConf is increased}. If the first access occurs outside this window, or an element leaves the SCS before being accessed, PatternConf decreases.

\prerebuttal{PatternConf is implemented as two 4-bit saturating counters per-PC, each behaving as above, but with different \textit{bias} factors. Rather than counting upwards and downwards symmetrically, the \textit{Base}PatternConf counts upwards by 1 and downwards by 2. This means it only reaches high values if the prefetch is useful more than 66\% of the time, rather than more than 50\% for a symmetric counter, which would permit inaccurate prefetches. The second (\textit{High}PatternConf) counts up by 1 and down by 5, to issue aggressive prefetches (\cref{ssec:aggression}) when we are more than 83\% ($\frac{5}{6}$) certain that prefetches are high quality. Both factors can be adjusted 
to alter the prefetcher's willingness to store metadata and issue prefetches, trading off coverage and performance for accuracy and traffic reduction.}

\subsubsection{Sampling Methodology}

We dynamically alter sampling rate for each PC via a 4-bit saturating counter stored in the training table, initialized to 8. We insert an entry into the History Sampler with probability:

\begin{gather*}
\dfrac{SamplerSize}{MaxSize} * 2^{SampleRate-8}
\end{gather*}

$MaxSize$ is the number of entries in a Markov table with maximum cache-partition allocation, and $SamplerSize$ is the (smaller) number of entries in the sampler. Entries placed into the sampler inevitably replace other, older entries that we then lose the opportunity to observe. If this happens too frequently, then we will fail to pick up long-term reuse patterns entirely. 

If the reuse distance of the victim entry is longer than \textit{MaxSize} then we are only replacing stale entries. We decrease the reuse confidence (if the element is unused) of the victim entry's PC, and increase the sampling rate of the replacement PC. Otherwise if the victim element is not older than MaxSize and also has not yet been used, we replace the potentially useful victim but reduce the sampling rate of the replacement's PC to reduce the probability of replacing useful data in the future. This allows us to ultimately see if repeat accesses exist for every PC: even if some PCs fill the cache more often than others, or reuse distances are very long, in both cases evicting data before we see repetition, we dynamically adjust fill rates of each PC to compensate so the sampler can store some elements from each to analyze all of them.

\subsection{Lookahead, Degree and Aggression Optimization}
\label{ssec:aggression}

Triage~\cite{Triage-MICRO19} observes that higher degrees, i.e. following chains of table entries to produce multiple prefetches per cache miss, give a significant increase in performance, by making prefetches more timely. However, accuracy drops and DRAM traffic grows dramatically between the highest-accuracy and highest-performance configurations (\cref{sec:eval}).

In \name{} we have more information about the likely success of our prefetches, since we have per-PC classifiers of both ReuseConf and PatternConf. It follows that for PCs where both of these are high, even high-degree prefetches will remain accurate. However, a high degree alone is insufficient to achieve good timeliness, as walking the linked Markov-table data structure may be as slow as the CPU's demand accesses for the same locations.
We can solve this problem by increasing the \textit{lookahead} to 2 from 1, instead of only increasing the degree. For an $(x,y,z,x,y,z)$ repetition, we store $(x,z)$, $(y,x)$ and $(z,y)$  instead of ($x,y)$, $(y,z)$ and $(z,x)$, to better hide latency via overlapping future prefetches.  %
Unlike increasing degree, increasing the lookahead distance requires changing the data stored in the Markov table itself. We must decide this at time of training rather than the time of metadata usage, and lookahead must be consistent (for an individual PC) over long periods of time to avoid targets being skipped. Unlike in the Markov table, which can still store a single payload, the training table must store a shift register as long as the largest lookahead possible: e.g. for the pattern $(x,y,z)$, if $x$ was written into lastAddr$[0]$ two accesses ago, then shifted it to lastAddr$[1]$ one access ago (overwriting lastAddr$[0]$ with $y$), we will see $z$ in the current iteration  and store $(x,z)$ in the table (followed by setting lastAddr$[0]$ to $z$ and shifting $y$ into lastAddr$[1]$ for the next training).

We set the lookahead distance to 2\footnote{\prerebuttal{While larger lookaheads can increase performance, they also require larger shift registers in the training table. Provided we have a lookahead of at least 2, we can use degree to get arbitrarily far ahead of the CPU. This is not true of lookahead-1 on a linked list for example, where the CPU will be able to walk the list as fast as the prefetcher if entries are cached in the L3, as the prefetcher has no more memory-level parallelism than the original program.}} (and thus the lookahead bit in the training table) if \prerebuttal{HighPatternConf} reaches its maximum value (15).  To avoid rapid switching, and thus entry skipping, we only return to lookahead 1 (which requires temporal locality only over one entry rather than two in order to be accurate, so it is better for uncertain patterns) if \prerebuttal{BasePatternConf} subsequently drops below its initial value (8).

We can then prefetch at higher degrees as well on top of the higher lookahead, to increase timeliness without increasing the training table's history shift register further in size. However, again we should only do this when we are confident in our prefetches, otherwise we should expect accuracy to become untenable (such as we see with the Triage at degree 4, \cref{sec:eval}). If \prerebuttal{HighPatternConf} is above the default 8, we issue chained lookups to the Markov table, generating up to 4 prefetches total.  When ReuseConf or \prerebuttal{BasePatternConf} are at their initial value \prerebuttal{(8, or half way)} or below, we neither issue prefetches nor store entries in the Markov table, to decrease L3 traffic and avoid storing useless metadata. %
\subsection{Metadata Reuse Buffer}
\label{ssec:metadatareusebuffer}

Increasing the degree causes redundant lookups in the L3 cache's Markov table. Triage's energy consumption doubles at degree 8~\cite{Triage-MICRO19}.
We add a small Metadata Reuse Buffer next to the prefetcher, storing the most recently used entries in the Markov table (along with the four set-index bits used to index the Markov table but not this smaller, 256-entry 2-way associative structure). %
 This is a 256-entry, 2-way set associative structure (so negligible in energy compared to a multi-megabyte highly associative L3 cache), and uses FIFO\footnote{\prerebuttal{We use FIFO rather than LRU here because this structure is too small for general temporal locality: rather, elements will be accessed four times then should leave the cache for other entries.}} replacement for accessed Markov entries that result in a prefetch. This removes the vast majority of redundant accesses to the L3, as any repeated accesses from overlapping walks caused by degrees higher than 1 will be eliminated. It also improves timeliness: repeat accesses will not incur an L3 latency penalty, causing most degree-4 (or higher) accesses to result in only a single L3-cache Markov lookup.
Triage's extra accesses to the L3 are not only caused by redundant accesses, i.e. repeats from one prefetch to the next, because in Triage's case many high-degree prefetches are inaccurate. But \name\ only increases the degree above 1 in scenarios where it is highly confident in the quality of its prefetches.

The Metadata Reuse Buffer allows one further optimization. When prefetches are accurate, the Markov entry $(x,y)$ we are about to update \prerebuttal{(on access to $y$)} will be in the buffer, because it has recently been used to generate a prefetch \prerebuttal{(on access to $x$}). If none of the information in the Markov table is due to change (the entry's target and confidence are identical to their previous values), we can avoid issuing the update to the L3.

\subsection{Set-Duelling Partition Sizing}
\label{ssec:metadatasize}

\begin{figure}
    \centering
    \includegraphics[width=\columnwidth]{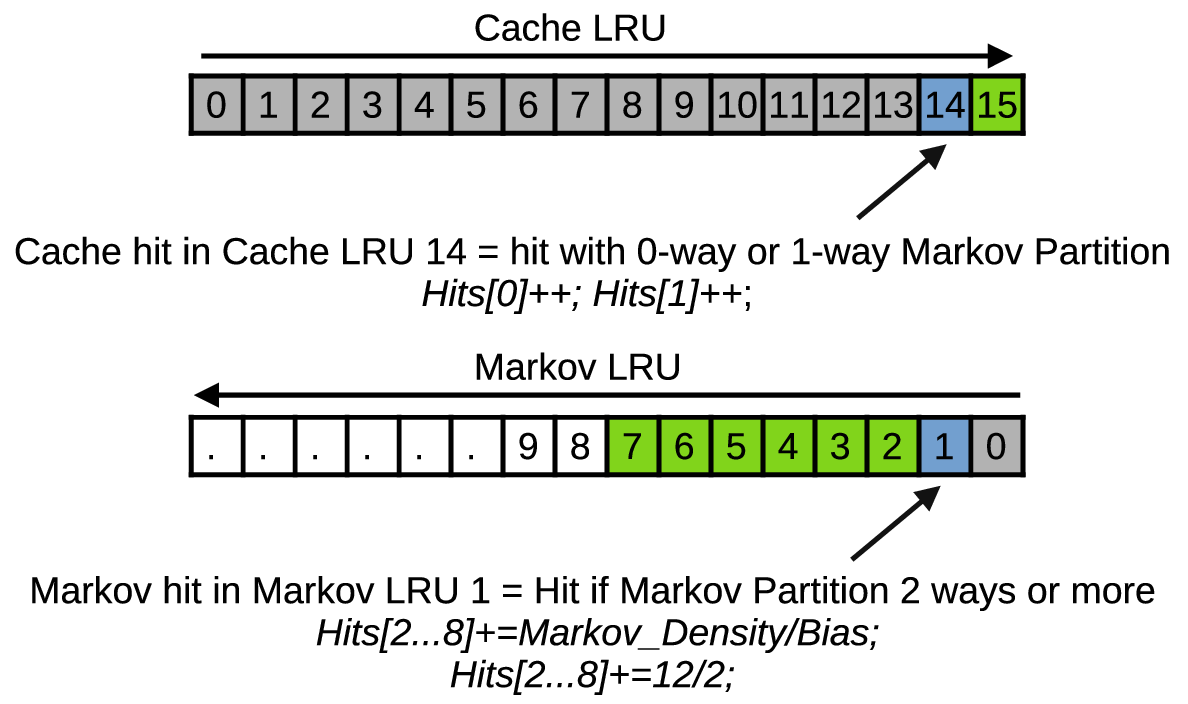}
    \caption{\prerebuttal{The Set Dueller samples 64 random sets and works out the best tradeoff between Markov table and Cache space, by modeling the cache hit rate of each of the 9 options for 8-way maximum partition in a 16-way cache, by keeping one copy of the Cache's tags and one copy of the Markov table's tags, each modeled as though they were LRU.}}
    
    \label{fig:setdueller}
\end{figure}

A bloom filter covering the full 196608 entries that can be stored in a 1MiB \name{} partition is infeasibly large (\cref{ssec:triagesizing}). \prerebuttal{While this can be cut down via sampling only a portion of the address space (and we model this for Triangel-Bloom in the evaluation, with an experimentally determined bias factor of $1.5$ which achieved the highest performance), a Bloom-filter policy still over-sizes the Markov table by prioritizing its entries over standard cache lines (\cref{ssec:triagesizing}).
} 

\prerebuttal{\name{} instead directly trades off cache hits versus Markov-table hits to achieve the highest hit rate, via a custom set-duelling~\cite{SetDuel} mechanism on 64 random cache sets.} We model two structures: a full-size L3 cache unaffected by Triangel's prefetching, and a full-size Markov table, and interpolate the partitioning between the two that gives the best hit rate.

\prerebuttal{The 64 sampled sets each store a single Markov-table tag and a single cache-line tag (compressed as 10-bit hash-tags) per cache way: 8 out of 16 cache ways can be allocated to Triangel in our case, so we only explicitly store 8 ways of Markov table, but 16 cache ways, modelled within the dueller as LRU\footnote{\prerebuttal{This is a very approximate model. Triangel does not use LRU replacement, and the cache might not either~\cite{Intel-RRIP}: we model LRU in the dueller because it gives unique evictability scores per-tag that allow us to model just two extremes (all-Markov and all-cache) and infer hit rates of the other 7 options. By contrast, RRIP~\cite{RRIP} may share the same evictability for multiple tags at once, and will insert in a different place depending on the associativity (an access of a newly inserted line might be in way 6 of a 7-way cache, or way 2 of a 3-way cache), meaning we cannot give each way a unique score. The modelled L3-cache hits are based on the miss/prefetch-hit streams seen by the prefetcher, rather than the true L3 state, to cancel out replacement-state changes by the prefetcher, and by partitioning.\ignore{Triangel is also always 12-way associative regardless of its partition size: the approximation that it instead has N-way-partition associativity lets us avoid storing all 96 Markov-table entries within each set.}}}. This gives each element in both tables a unique eviction priority, indicating whether they would hit in each connfiguration, from a Markov-table reservation of 0 to 8 ways. We also store 9 32-bit counters (total, not per sampled set), one for each possible partitioning decision. As shown in \cref{fig:setdueller}, when the $ith$ most evictable cache element is accessed in a sampled set, counters are incremented corresponding to partitions where this access would be a hit: where $i$ or more ways are used as data cache. Likewise, each time the $ith$ most evictable Markov-table entry is accessed in a sampled set, counters corresponding to partitions where $i$ or more ways are reserved for the Markov table are incremented.}\footnote{
\prerebuttal{This is made harder by Markov-table entries and cache lines being different sizes, since 12 Markov-table entries fit in a single cache line, so Markov-table entries' lifetimes in the cache are much longer than cache lines', and the indexing policy is very different as a result (\cref{ssec:index}). We handle this by sampling 1/12 as many entries of the Markov table, to make the ratios of the two match, but treating each sampled Markov-table hit as being worth 12 cache-line hits. However, since prefetches cause DRAM accesses whereas cache hits do not, we also bias against Markov entries by a factor \textit{B} (2 by default, so each Markov hit is worth 6 cache-line hits).}  %
}

\prerebuttal{We use a 500000-entry window to track these counters, after which they are reset and the optimal partition from the last window used to set the partition for the next. This is a tradeoff between reactiveness to new workload phases versus limiting expensive re-indexing (\cref{ssec:metadataformat}), though resizes are rare and this parameter is not sensitive outside extreme values. }

\subsection{Sizing}
\label{ssec:sizing}

\begin{table}[t]
\begin{center}
    \begin{tabular}{c|c|c}
 Table & Entries & Size  \\
 \hline
Training Table & 512 & 7808B \\
 History Sampler & 512 & 6080B\\
 Second-Chance Sampler & 64 & 584B \\
Metadata Reuse Buffer & 256 & 1472B \\
  \prerebuttal{Set Dueler} & 64$\times$(8+16) & 2106B \\
 \hline
 \prerebuttal{Total} & & 17.6KiB
\end{tabular}
\end{center}

\caption{Sizing of Triangel's structures.}

\label{tab:sizetable}
\end{table}

\name{} adds \prerebuttal{four} structures to Triage: the Reuse Buffer, the History- and Second-Chance Samplers\prerebuttal{, and the Set Dueller}. It increases the size of each training-table entry (\cref{tab:training}) by 76 bits: 32 to increase the lookahead distance (LastAddr$[1]$ and Lookahead), and 44 for history sampling (Timestamp, PatternConf and SampleRate). By removing the 1024-entry upper-tag lookup table (3840B), HawkEye's Dueller (13KiB~\cite{HawkEye}) by using SRRIP~\cite{RRIP} instead, and the Bloom Filter (200KiB for 5 percent error), we save significantly: \name{} has \prerebuttal{17.6KiB} of total dedicated storage (\cref{tab:setuptable}), versus 219.5KiB for Triage.

%% file: sections/setup.tex
\section{Experimental Setup}

\begin{table}
\begin{tabularx}{\columnwidth}{lX}
Core & 5-Wide, out-of-order, 2GHz \\
Pipeline & 288-Entry ROB, 120-entry IQ, 85-entry LQ, 90-entry SQ, 150 Int / 256 FP registers, 4 Int ALUs, 2 Mul/Div, 4 FP/SIMD, 2 R/W Ports \\
Branch &  64KiB MPP-TAGE \\
L1 ICache & 64KiB, 4-way, 2-cycle hit lat, 16 MSHRs \\
L1 DCache & 64KiB, 4-way, 4-cycle hit lat, 16 MSHRs, deg-8 stride pf \\
L2 Cache & 512KiB, 8-way, 9-cycle hit lat, 32 MSHRs \\
L3 Cache & 2MiB/core, 16-way, 20-cycle hit lat, 36 MSHRs \\
Memory & LPDDR5\_5500\_1x16\_BG\_BL32\\
OS & Ubuntu 22.04 \vspace{4pt}
\end{tabularx}
\vspace{-8pt}
\caption{Core and memory experimental setup.}
\vspace{-15pt}
\label{tab:setuptable}
\end{table}

We evaluate \name{} using gem5 v23.0.0.1, with the configuration in \cref{tab:setuptable}. The cache configuration is set up to be a close match to the original Triage~\cite{Triage-MICRO19} paper, with parameters otherwise chosen to be a close match for the Arm Cortex X2~\cite{X2-SOG,Wikichip-X1,Anandtech-X2}. Gem5 is run in Full-System mode, with KVM used to generate checkpoints from throughout the execution of each workload. Each experiment consists of 20 checkpoints per workload, each warmed up for 50000000 instructions and sampled for 5000000 instructions. \prerebuttal{Since it impacts so many other metrics, we also show a breakdown of Triangel with a Bloom filter instead of the default Set Dueller is every graph, with other parameter tuning shown separately.}

We implement Triage based on the details given in the two papers~\cite{Triage-MICRO19,TriageISR-ToC22}, with ambiguous components matching the public implementation~\cite{Triage-Codebase} and/or cleared up in \cref{sec:inconsistencies}. Both Triage and \name{} allow their Markov tables to use half the last-level cache maximum. Each Markov-table access has a 25-cycle hit latency, accounting for 20 cycles of L3 cycle access time    and 5 cycles of compressed-metadata handling. Both prefetch into the L2, and use the same sized structures where relevant (\cref{ssec:sizing}). Triage uses HawkEye~\cite{HawkEye} replacement and \name{} the simpler SRRIP~\cite{RRIP}. For Triage, we evaluate both their degree-1 default setup (\textit{Triage}), and their faster but less accurate degree-4 setup (\textit{Triage-Deg4}). To isolate Triangel's aggression control from its other improvements, we further add Triangel's new lookahead mechanism to Triage (\textit{Triage-Deg4-Look2}).

To maintain a close evaluation to the original papers~\cite{Triage-MICRO19,TriageISR-ToC22}, experiments use the 7 most irregular, memory-intensive workloads from SPEC CPU2006~\cite{SPEC2006}: Xalancbmk, Omnetpp, Mcf, GCC (166 input), Astar, Soplex (3500 ref.mps input) and Sphinx3, on the Ref inputs. 
Triage does not reduce performance on the remainder of SPEC CPU2006~\cite{Triage-MICRO19}, and \name{} is less willing to prefetch or take up L3 space when it is not useful, and so we avoid repeating these experiments. 
\prerebuttal{We also evaluate on multiprogrammed combinations of these workloads (where Triangel's structures are core-private save for the Markov partition and its Set Dueller, which are shared) to show a more bandwidth-constrained environment, and evaluate on Graph500 Search~\cite{murphy2010introducing} (s16 e10 7MiB input and s21 e10 700MiB input~\cite{1204_Ainsworth,1256_Ainsworth}) to evaluate performance when a workload is cache- and memory-intensive but with neither temporal correlation nor a small enough working set for temporal prefetching, to stress the techniques further.}

%% file: sections/evaluation.tex
\section{Evaluation} \label{sec:eval}

\begin{figure}[t]
    \centering
    \includegraphics[width=\columnwidth]{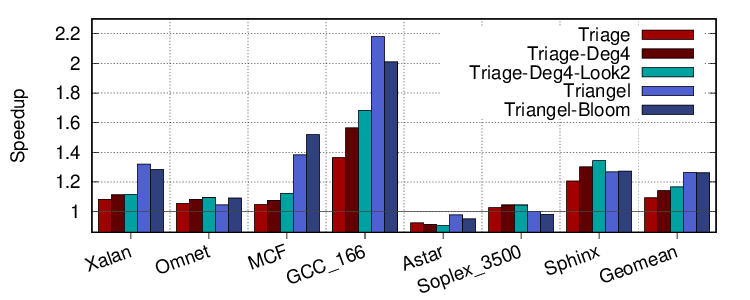}
    \caption{Speedup}
    \label{fig:speedup}
\end{figure}

\begin{figure}[t]
    \centering
    \includegraphics[width=\columnwidth]{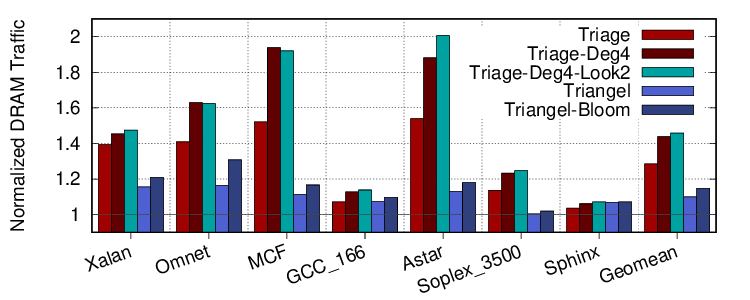}
    \caption{Normalized DRAM Traffic (lower is better).}
    \label{fig:memreadtraffic}
\end{figure}

\begin{figure}[t]
    \centering
    \includegraphics[width=\columnwidth]{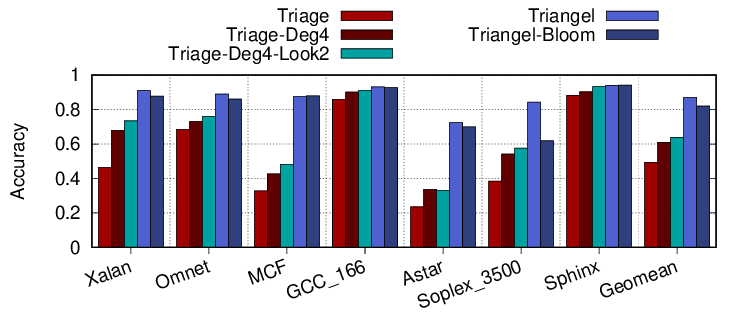}
    \caption{Accuracy (prefetched lines used before L2 eviction).}
    \label{fig:accuracy}
\end{figure}

\begin{figure}[t]
    \centering
    \includegraphics[width=\columnwidth]{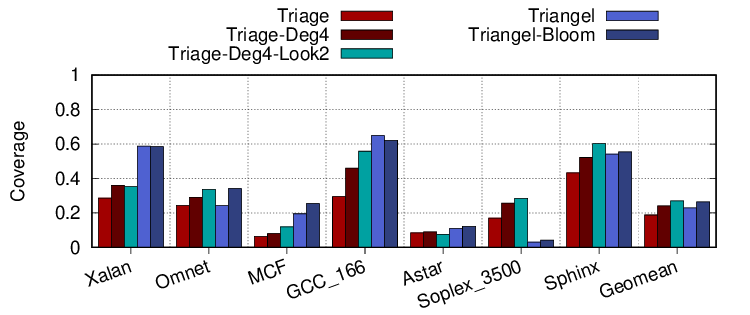}
    \caption{Coverage (CPU L2-cache demand misses 
   eliminated from baseline).}
    \label{fig:coverage}
\end{figure}

\begin{figure}[t]
    \centering
    \includegraphics[width=\columnwidth]{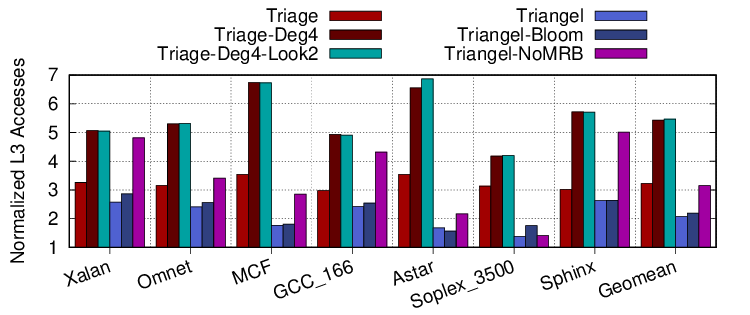}
    \caption{\prerebuttal{Normalized L3-cache traffic, including Markov-table accesses and L3 data accesses (lower is better).}}
    \label{fig:l3traffic}
\end{figure}

\Cref{fig:speedup} shows that, on the workloads examined in the original papers on Triage~\cite{Triage-MICRO19,TriageISR-ToC22}, \name{} achieves a speedup of \speeduptriangel{} geomean, versus \speeduptriage{} for Triage. \prerebuttal{\Cref{fig:memreadtraffic} shows this is despite inducing significantly less memory traffic: only 10\% above baseline, versus 28\% for Triage, including overheads from taking resources away from the L3 cache.}

\subsection{Analysis}

The performance improvement of \name{} is primarily from its increased aggression (degree-4 and lookahead-2 when the history sampler is confident) allowing it to significantly improve timeliness, but in spite of this, \name{} is significantly accurate than Triage (\cref{fig:accuracy}). Curiously, Triage-Deg4 is also more accurate than Deg-1 by ratio, but the sheer volume of incorrect prefetches created causes traffic issues; Triage-Deg4 (same maximum aggression as Triangel) achieves only 14.2\% speedup, since it cannot selectively target high-quality streams, nor use Triangel's high-lookahead (\cref{ssec:aggression} or metadata-reuse (\cref{ssec:metadatareusebuffer}) mechanisms to improve timeliness. Adding Triangel's Lookahead-2 to Triage-Deg4 improves things but only slightly (16.6\%); Triangel's aggression controls and metadata filtering and formatting are needed to gain maximal benefit. %

\prerebuttal{\name{} achieves only a slightly higher performance with its default Set Dueller  (\speeduptriangel{}) than with a Bloom Filter (26.1\%), but achieves significantly lower DRAM traffic (10\% versus 14.6\%). 
We consider the tradeoffs in \cref{ssec:ablation}. %
}

\prerebuttal{Coverage (\cref{fig:coverage}) shows a more complex picture. Both Triangels are less willing to prefetch from poor-quality streams such as Astar and Soplex -- whereas Omnet and MCF see lower coverage on the Set-Duelled Triangel versus the Bloom-filter Triangel in order to alleviate DRAM requests. %
}

\begin{figure}
    \centering
    \includegraphics[width=\columnwidth]{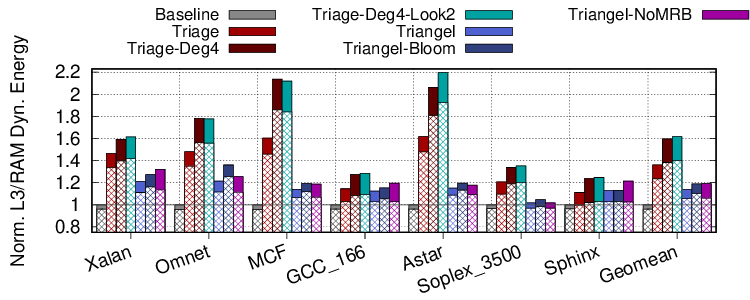}
    \caption{Normalized combined DRAM+L3 dynamic energy (lower better). \prerebuttal{Hashed bars represent the DRAM proportion.}}
    \label{fig:l3dynen}
\end{figure}

\subsection{Energy Consumption}

Because \name{} discards many poor-quality PCs that do not show history patterns, it accesses the Markov table significantly less than Triage in every workload, significantly reducing the traffic to the L3 cache (\cref{fig:l3traffic}). This is in spite of the much higher prefetch degree in \name{}, where the \prerebuttal{Metadata Reuse Buffer (versus NoMRB) successfully prevents repeat accesses} to the same locations even as they are re-walked over overlapping high-degree prefetches, unlike in Triage-Deg4 which exceeds $5\times$ the number of accesses. Indeed, improved accuracy means that even without an MRB, Triangel (Deg-4) still only causes as many L3 accesses as Deg-1 Triage, which itself does not benefit from an MRB due to not issuing high-degree prefetches.

\name 's lower number of memory accesses, and lower number of Markov-table accesses, provide significant energy savings. To estimate this, we use the same methodology as in the original Triage paper~\cite{Triage-MICRO19}: we assign DRAM accesses an energy cost of 25 units, and L3 accesses (including data accesses and Markov-table accesses) a cost of one unit. We then compare this against the number of DRAM accesses and L3 accesses in the baseline (which has no Markov-table accesses and no temporal prefetcher). \prerebuttal{In \cref{fig:l3dynen}, \name{} (14\%) is significantly lower than both Triage (36\%) and Triage-Deg4 (60\%) -- and, due to the Set Dueller's metadata-sizing tradeoffs, Triangel-Bloom (19\%).} 

\subsection{Multiprogrammed Workloads}

\begin{figure}[t]
    \centering
    \includegraphics[width=\columnwidth]{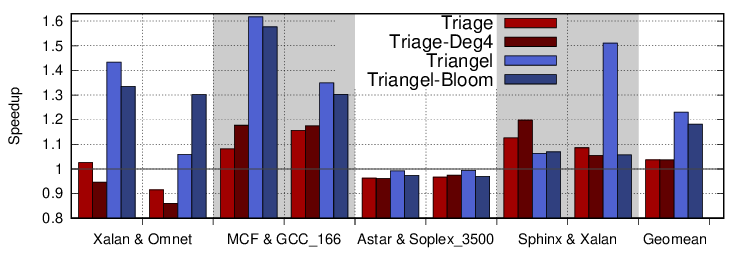} 
    \caption{\prerebuttal{Multiprogrammed-workload speedup. Workloads paired and prefetched simultaneously on two cores.}}
    \label{fig:multiprogram}
\end{figure}

\prerebuttal{\Cref{fig:multiprogram} shows the same workloads run in adjacent pairs on two cores simultaneously (with Xalan doubled to make an even set). 
Triangel typically maintains its performance, while Triage slips further behind. Triage-Deg4 suffers new slowdowns, not exceeding Triage in geomean, because its aggression is particularly misplaced when bandwidth-constrained.}

\subsection{Adversarial Workloads}

\begin{figure}[t]
    \centering
    \includegraphics[width=.49\columnwidth]{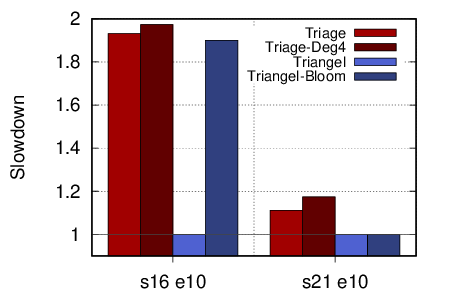} \includegraphics[width=.49\columnwidth]{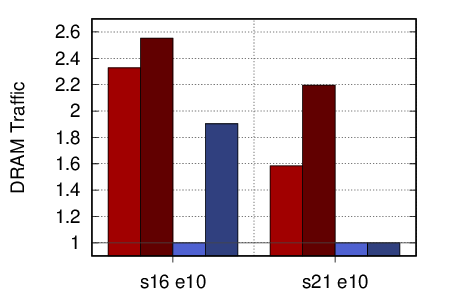} 
    \caption{\prerebuttal{\textit{Slowdown} and DRAM traffic for Graph500 search.}}
    \label{fig:graph500}
\end{figure}

\prerebuttal{\Cref{fig:graph500} shows the slowdown and DRAM traffic for a workload unsuited to temporal prefetching: Graph500 search~\cite{murphy2010introducing}. Neither input exhibits temporal correlations: the pattern for the 700MiB s21 e10 graph is too large for a prefetcher with maximum capacity 12MiB (16MiB for Triage), and while the 7MiB s16 e10 can fit, its accesses show too little repetition for prefetches to be worthwhile. Triage has no concept of accuracy and will always increase to maximum capacity given large enough input. This means the Triage techniques lower performance dramatically, since they maximize Markov-partition size and minimize L3 hits, while generating inaccurate prefetches. Triangel-Bloom also does poorly on s16 e10, even though it rarely generates prefetches: this is because its PatternConf counters transiently lock on to small patterns of genuine sequences, which are enough to fill the Bloom filter, but do not last long enough to keep PatternConf above threshold. Standard Triangel's Set Dueller realises that prefetch hits are rarer than cache hits. For s21 e10, neither Triangel activates: the reuse distance is beyond ReuseConf.}

\subsection{Markov-Table Format: Analysis on Triage}
\label{ssec:triageanalysis}

\begin{figure}[t]
    \centering
    \includegraphics[width=\columnwidth]{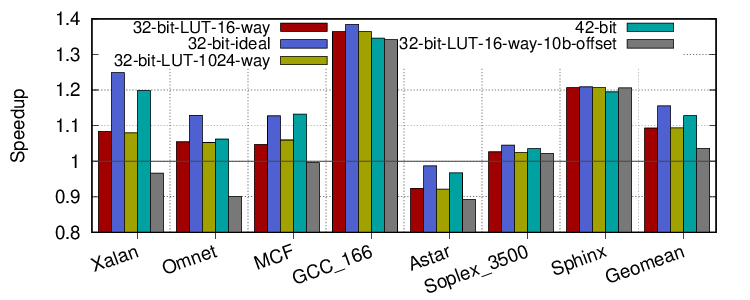}
    \caption{Performance of Triage with different Markov-table entry sizes, and Lookup-Table configurations in cases where the Markov-table entry is 32-bits and so requires lookup to reconstruct a full prefetch target (first is default).}
    \label{fig:triagelut}
\end{figure}

In \cref{fig:triagelut} we justify our Markov-table format in \name{} by evaluation of Triage under various scenarios. The first (32-bit-LUT-16-way, the configuration used elsewhere) stores metadata as described in \cref{ssec:metadataformat}, with each entry 32-bits wide, using a 1024-entry lookup table to generate the full target address~\cite{Triage-MICRO19}. Though the associativity of the lookup table is not given in the original paper, we see that a 16-way lookup table performs no worse than a fully associative lookup table (32-bit-LUT-1024-way). However, there is a significant drop from a hypothetical mechanism with a perfect lookup table (32-bit ideal). There is also a performance increase from the simpler strategy of removing the lookup table, and storing each entry as 42-bits long directly (\cref{ssec:markov}, 42-bit) as in \name{}. Minor changes in accesses cause even worse behavior. If we change the lookup table to cover one more bit, meaning the offset stored is 10 bits long instead of 11 as in the original paper (32-bit-LUT-16-way-10b-offset), roughly equivalent to halving physical-page locality or doubling page fragmentation, performance drops dramatically. 

\begin{figure}[t]
    \centering
    \includegraphics[width=\columnwidth]{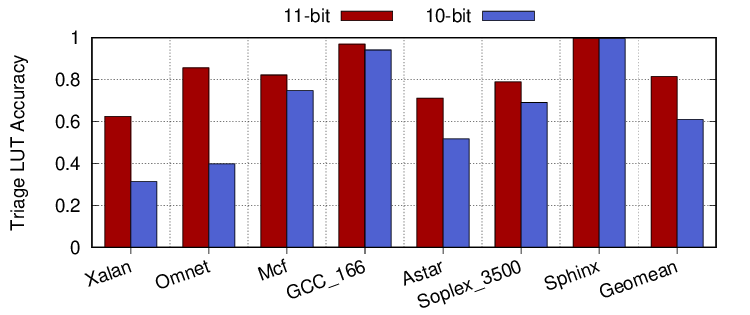}
    \caption{Accuracy of Triage's LUT with varying number of offset bits (11-bit default). When capacity is exhausted, we see wrong prefetches. \name{} avoids its use.}
    \label{fig:triagelutacc}
\end{figure}

We see in \cref{fig:triagelutacc} that, while the LUT works well for GCC and Sphinx, accuracy is poor for others and plunges with fragmentation modeled by 10-bit offsets. Unlike the Markov table, which stops generating prefetches if its capacity is exhausted, the lookup table (accessed only via index) returns addresses the program may never have accessed. It is sensitive to minor changes in input, and assumes locality possible only on a freshly booted system. By removing the LUT, \name{} is immune from assumptions about physical-frame locality.

\subsection{Ablation Study}\label{ssec:ablation}

\begin{figure}[t]
   \centering
       \subfloat[Speedup]{
    \includegraphics[width=\columnwidth]{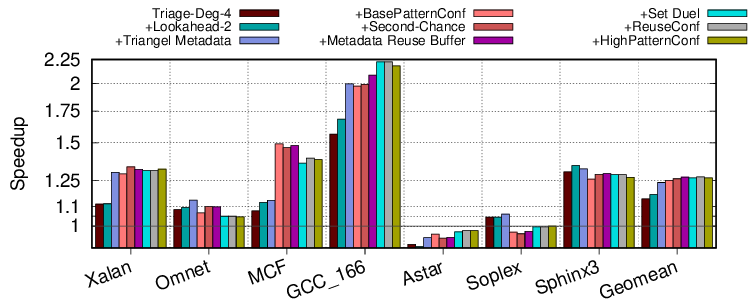}
    }
    
        \subfloat[DRAM Traffic]{
    \includegraphics[width=\columnwidth]{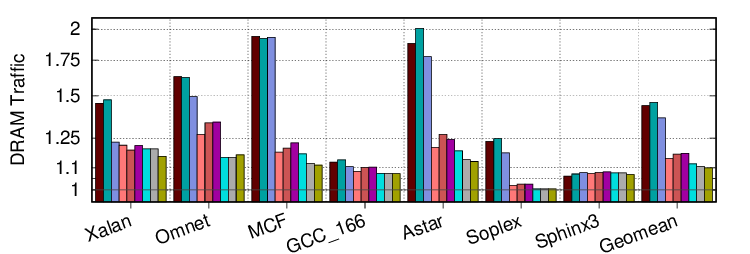}
    }
\caption{\prerebuttal{Impact of progressively adding individual features to form Triangel, starting with Triage Degree 4.} }
\vspace{-10pt}
    \label{fig:speedupsmall}
\end{figure}

\Cref{fig:speedupsmall} shows the contribution of each individual change that Triangel introduces, starting with Triage Degree-4 and progressively adding new mechanisms.

Adding Triangel's Lookahead-2 mechanism (\cref{ssec:aggression}) improves performance by improving timeliness, but only slightly; the high inaccuracy results in overwhelming DRAM traffic and limited benefit. The switch to Triangel's Markov table format removes the inaccuracy of Triage's lookup table (\cref{ssec:triageanalysis}), which compounds over high degrees to generate progressively more incorrect prefetches.

The use of BasePatternConf, to prevent storage of prefetch metadata, and generation of prefetches, for patterns with less than 66 percent accuracy, halves the DRAM traffic overhead. It results in a substantial performance improvement on MCF, which can now use its limited metadata storage for more profitable patterns. However, it also makes Omnet and Sphinx3 too conservative, as these workloads have strong temporal reuse but not always in strict sequence (so prefetches still get used before they are evicted from the cache). The Second-chance sampler's ability to track non-strict-sequential patterns mitigates the performance impact.

The Metadata Reuse Buffer (\cref{ssec:metadatareusebuffer})'s impact on performance is slight except for on GCC; Triangel is already timely due to it's high degree and lookahead. However, we previously saw it has a favorable impact on energy (\cref{fig:l3dynen}).

The Set Dueller (\cref{fig:setdueller}) significantly reduces DRAM traffic by directly considering the tradeoff between the cache being used for Markov-table storage versus data storage. In particular, it speeds up GCC through reducing traffic, at the expense of slowing down Omnet and MCF, where the default values for the Set Dueller decide the performance improvement is not worth the extra traffic (though more aggressive tradeoff parameters, not shown here, do increase performance).

With the exception of Astar and MCF, all of the workloads here have working sets small enough to not trigger ReuseConf, though MCF in particular sees speedup by ReuseConf not wasting storage on patterns too large to fit in the L3. Finally HighPatternConf lowers DRAM traffic and performance, by triggering high-degree and high-offset prefetching only when confidence reaches the higher $\frac{5}{6}$ threshold rather than BasePatternConf's $\frac{2}{3}$ threshold for storing metadata, deliberately making the prefetcher less aggressive, though typically our workloads' access patterns fall above or below this range.

%% file: sections/related.tex
\section{Related Work}

Surveys of prefetchers have been performed by both Falsafi and Wenisch~\cite{1197_Babak}, and Mittal~\cite{1198_Mittal}. We discuss the most relevant work, on correlating prefetchers, below.

\subsection{Markov Prefetchers}

Markov prefetchers~\cite{Markov}, like Triage~\cite{Triage-MICRO19} and \name, store address pairs, for correlation prefetching. While such tables were originally proposed with multiple successors (width $>$1)~\cite{Markov}, Triage and \name{} store a single successor, to improve density and accuracy at the expense of coverage. To increase timeliness, several approaches have been explored: increased degree by walking multiple entries~\cite{DynHot}, storing streams of successors~\cite{UserLevel} rather than single elements in each table entry (at the expense of density), and storing groups of prefetches~\cite{Epoch} within an epoch at the index of the first miss (used by Triage-ISR~\cite{TriageISR-ToC22} for contiguous regions). \name{} also increases the lookahead of the prefetcher (to prefetch at high offsets, common in other types of prefetcher~\cite{BOP,1204_Ainsworth,1256_Ainsworth}).%

\subsection{Aggression Control}

One of \name's primary tasks is controlling the aggression of Triage via the History Sampler. This is made harder than the traditional problem of controlling prefetch aggression, because decisions need to be made as to whether to \textit{store} Markov-table metadata in the first place, and entries are stored long before they are used, requiring long-term analysis. %
For the decision on whether to \textit{issue} a prefetch, Srinath et al.~\cite{Feedback} track global prefetcher pollution and lateness using tag bits, and use this to alter distance, degree, and LRU state on prefetch fill. PACman~\cite{DBLP:conf/micro/WuJMSE11} uses set duelling to for rereference prediction of prefetches to influence replacement state. Ebrahami et al.~\cite{10.1145/1669112.1669154} coordinate prefetchers in multicores to reduce interference. Perceptron-based Prefetch Filtering (PPF)~\cite{PerceptronPrefetchFilt} handles a similar task to \name{}, of improving aggression while maintaining accuracy using classifiers. Unlike \name{}, it classifies over many more properties than just the PC, but the large state space and long reuse distances in temporal prefetching makes this a challenge relative to the prefetchers in PPF.

\subsection{Other Address-Correlating Prefetchers}

Triage chose a Markov-table as the structure for its address-correlating metadata because of its relative density. For prefetchers with off-chip metadata storage, structures that trade storage cost in favor of timeliness are common. Global history buffers~\cite{GHB,TemporalStreaming} do this through a layer of indirection: all missing addresses are stored in a single, linear buffer. Upon a miss, the history buffer entry is found via lookup in the index table. Subsequent misses are spatially local in the history buffer, so arbitrary degrees of prefetching can be achieved without chained lookups. Wenisch et al~\cite{PracticalMetadata} reduce the cost of updating the off-chip, non-spatially-local index table via sampling to only fill in the table on a fraction of addresses (each of which then generates several prefetches).

Irregular Stream Buffers~\cite{ISB} use a similar two-layer metadata format, based on assigning cache blocks a \textit{structural address} that is sequential based on miss patterns. The history buffer is replaced by a structural-to-physical mapping table, and prefetches occur by translation of lookup address via physical-to-structural table, which gives the relevant location of a stream of successors in the structural-to-physical table. This reduces storage cost compared with GHBs by enforcing that every address is only in each table once. MISB~\cite{MISC-ISCA19} stores per-PC irregular stream buffers in-memory, removes the earlier constraint~\cite{ISB} that physical-to-structural table entries are brought in and evicted based on TLB entries, and introduces metadata prefetching techniques, though sees high DRAM-traffic overheads~\cite{Triage-MICRO19}. Voyager~\cite{10.1145/3445814.3446752} uses an online-trained LSTM neural network to improve the accuracy of address correlations though needs infeasible computation and storage.

\subsection{Non-Address-Based Correlating Prefetchers}

Tag-Correlating prefetchers~\cite{TagCorrelating} reduce table sizes by assuming the same relationship between the same tag at different indices, and that the index of both Markov-table entry pairs will match. Delta-correlating prefetchers (such as GHB/DC as opposed to GHB/AC, both from the same paper~\cite{GHB}) correlate the differences between addresses rather than the absolute addresses themselves. When deltas are used for both indices and targets, the resulting table can be smaller (as different addresses reuse the same entries if they have the same deltas) and spatial patterns can be picked up allowing coverage of patterns otherwise the remit of stride prefetchers~\cite{1194_Chen}, at the expense of losing the ability to correlate absolute addresses, and aliasing different address pairs with the same deltas as each other, hampering temporal-prefetching ability.

%% file: sections/conclusion.tex
\section{Conclusion}

Temporal prefetching has recently moved from the theoretical to the practical. Here we have described and fixed the inconsistencies in the state-of-the-art Triage prefetcher~\cite{Triage-MICRO19,TriageISR-ToC22} to derive configurations suitable for real-world deployment. We have significantly improved upon it through \name{}, using new structures to sample the miss stream to pick up and record high-quality, accurate prefetches, and issue them aggressively and efficiently when they will be effective.

We hope that this is the start of a wide set of new work in temporal prefetching. We have released our implementations of both techniques as open-source extensions to gem5 to accelerate the development of this vital new technology.

%% file: sections/artifact.tex
\section*{Artifact Appendix}

\subsection{Abstract}

Our artifact contains a modified gem5 simulator implementing both the Triangel temporal prefetcher from ``Triangel: A High-Performance, Accurate, Timely On-Chip Temporal Prefetcher'', ISCA 2024, as well as an implementation of the Triage~\cite{Triage-MICRO19,TriageISR-ToC22} prefetcher (MICRO 2019) for comparison against.

It also contains scripts to run the seven SPEC-CPU 2006 benchmarks evaluated in the paper, and instructions on how to generate checkpoints from throughout the programs' executions by using KVM within gem5 in full-system mode.

\subsection{Artifact check-list (meta-information)}
{\small
\begin{itemize}
  \item {\bf Algorithm:} Triangel prefetcher
  \item {\bf Program:}  SPEC CPU2006 (not supplied)
  \item {\bf Run-time environment:} Linux to run the gem5 simulator on (we used Ubuntu).
  \item {\bf Hardware:} An x86-64 machine with sudo access (to install dependencies and mount images).
  \item {\bf Metrics:} Speedup, DRAM Traffic, Accuracy, Coverage
  \item {\bf Experiments:} Baseline, Triage prefetcher, Triage Degree-4 prefetcher, Triangel prefetcher, Triangel-Bloom prefetcher
  \item {\bf How much disk space required (approximately)?: } 50GB
  \item {\bf How much time is needed to prepare workflow (approximately)?: } Around 1 hour to compile gem5. Around 30 minutes each to generate checkpoints for the seven workloads, if not using prebuilt checkpoints.
  \item {\bf How much time is needed to complete experiments (approximately)?: } Around 5 hours (if running all benchmarks in parallel on 20 checkpoints each), one hour per configuration.
  \item {\bf Publicly available?: } Yes
  \item {\bf Code licenses (if publicly available)?:} gem5 license
  \item {\bf Archived (provide DOI)?: } \url{https://doi.org/10.5281/zenodo.10892184}
\end{itemize}
}

\subsection{Description}

\subsubsection{How to access}
Clone the git repository at 
	
\begin{lstlisting}
git clone https://github.com/SamAinsworth/gem5-triangel
\end{lstlisting}

\subsubsection{Hardware dependencies}

Any recent x86-64 system running Ubuntu should suffice. Other Linux or Mac operating systems may also work (or Windows under WSL), perhaps with altered package dependencies, but are untested.

\subsubsection{Software dependencies}

Our simulator requires several package dependencies, which can be automatically installed by our scripts (scripts/dependencies.sh). To regenerate checkpoints for benchmarks from scratch, you will need access to a SPEC CPU2006 .iso file, placed in the root of the repository.

\subsection{Installation}

You can install this repository as follows:

\begin{lstlisting}
git clone https://github.com/SamAinsworth/gem5-triangel
\end{lstlisting}

All scripts from here onwards are assumed to be run from the \texttt{run\_scripts} directory, from the root of the repository:

\begin{lstlisting}
cd gem5-triangel
cd run_scripts
\end{lstlisting}

To install software package dependencies, run

\begin{lstlisting}
./dependencies.sh
\end{lstlisting}

Then, to compile gem5, run

\begin{lstlisting}
./build.sh
\end{lstlisting}

If you are generating checkpoints from scratch using KVM, follow the instructions in the README.md file.

Checkpoints should be in a folder, per-benchmark, inside the ``Checkpoints'' folder at the root of the repository (e.g. Xalan checkpoints are stored inside \texttt{gem5-triangel/Checkpoints/Xalan/m5out/cpt$*$}).

Your Ubuntu image, for gem5 to access in FS mode, should be in the root of the directory, as \texttt{x86-ubuntu} (generated as in \textit{Generating Your Own Checkpoints} in README.md).
\subsection{Experiment workflow}

Inside the \texttt{run\_scripts} folder, run

\begin{lstlisting}
./run_experiments.sh
\end{lstlisting}

This will run experiments for all folders inside \texttt{Checkpoints}.

If any unexpected behavior occurs, try removing the ``\&" inside the run\_experiments.sh to run the workloads sequentially rather than in parallel, to observe the errors, and if no obvious solution becomes apparent, please contact the authors.

\subsection{Evaluation and expected results}

Once your experiments are finished, and again inside the \texttt{run\_scripts} folder, run

\begin{lstlisting}
./analyse_experiments.sh
\end{lstlisting}

This will print various metrics to the terminal. If you are using our exact checkpoints, these should match the ones in \texttt{EXAMPLE\_RESULTS.txt} in the root directory. If you are using your own checkpoints of the same workloads, the trends should be comparable but the results will not be identical due to different sampling.

\subsection{Experiment customization}
The prefetcher itself is implemented inside src/mem/cache/prefetch/ -- see Triangel.cc and Triage.cc. See Prefetcher.py in the same folder for the various options available -- and in configs/common/CacheConfig.py to see how they are connected. configs/common/Options.py shows the options available on the command line. We also modified the cache system to allow reserving part of the L3 for prefetch metadata, to fix cross-page prefetching in gem5, and to model access-time delay to the Markov table -- see triangel.cc and modifications to the cache structures in the commit history for more details.

You can also run on your own checkpoints if you follow the guide in the github readme -- or more generally on other workloads by specifying both \texttt{--triangel} and \texttt{--p2sl3cache} (the latter to give cores a private L2 and shared L3 cache -- where by default Triangel attaches to the former and uses storage in the latter).

\subsection{Methodology}

Submission, reviewing and badging methodology:

\begin{itemize}
  \item \url{https://www.acm.org/publications/policies/artifact-review-badging}
  \item \url{http://cTuning.org/ae/submission-20201122.html}
  \item \url{http://cTuning.org/ae/reviewing-20201122.html}
\end{itemize}